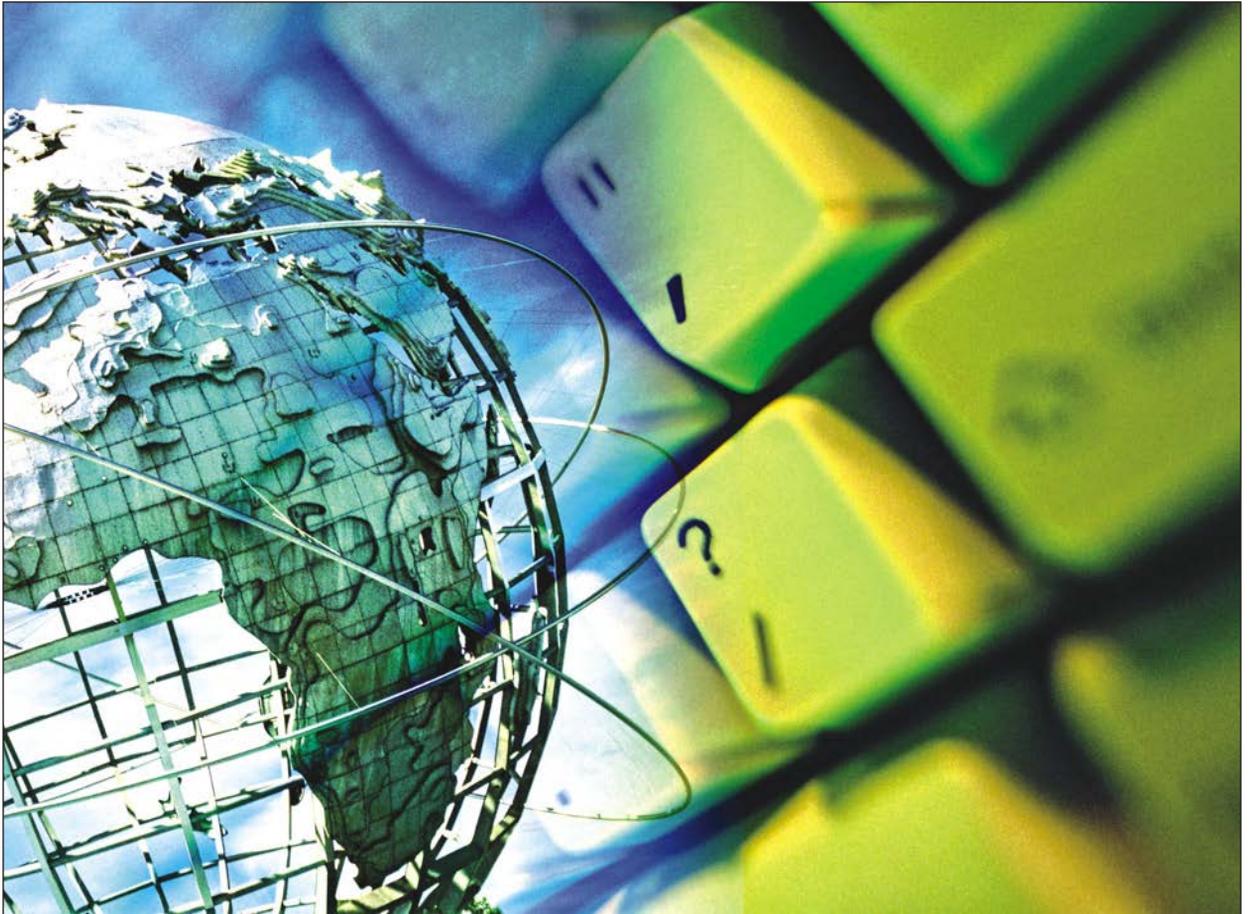

# Research Journal of
# Information Technology

## Research Article
## A Road Map to Bio-inspired Software Engineering

Said Ghoul

Laboratory of Bio-inspired Systems Modeling Research, Department of Computer Science, Philadelphia University, P.O. Box 1, 19392 Amman, Jordan

## Abstract

**Background:** Software production study is rapidly evolving in two parallel approaches: Conventional and bio-inspired. Bio-inspired approaches are generally developed and presented as enhancements of conventional ones. However, conventional approaches benefit from their integration with their global context, through software engineering methodologies for being advantageous. **Materials and Methods:** The integration of bio-inspired approaches, with bio-inspired software engineering methodologies will enrich bio-inspired approaches and let them be irrefutably the best. This study identifies the motivations of the emergence of such bio-inspired software engineering, presents a first approach to it with a road map and some of its challenges. **Results:** The application of this first approach on different software systems categories is presented with its summary evaluation. The richness and expressiveness of the concepts introduced by bio-inspired methodologies are strong compared with the conventional ones. However, the evaluation on real industrial software scale remains an open challenge. **Conclusion:** The obtained results prove the power, the effectiveness and simplicity of the bio-inspired methodologies compared to the conventional ones. The conventional software engineering is not inspired from nature processes and therefore, there is a gap between his concepts and mechanisms and those of real world. This leads to complexity and poverty in its models and their applications. However, this is the strength of the bio-inspired software engineering.

## INTRODUCTION

Software engineering is a technology allowing the production of software from user requirements[1]. A deep gulf separates these requirements, which are at a very high abstraction level (natural) from software product, which is practically at the low abstraction level. The secure move, through this deep gulf is guided by methodologies. Each methodology is defined by a set of steps organized and controlled by coordination rules. Each step carries out an activity. Thus, a methodology is a motherboard of software engineering, supporting coherent integration of techniques that collaborate to produce a software product. Software methodology has a determinant impact on product quality and cost. The active recent studies[2-7] in software engineering, show clearly some important challenges related to: Methodologies, cost and quality, abstractions levels, continuity, variability, autonomy and automation etc.

Biology, being rich with a variety of high quality production systems has inspired researchers with alternative techniques in order to face the above challenges. So, bio-inspired techniques emerged and have been in a rapid development in the last two decades. The obtained results were stimulant and lead to development of bio-inspired methods supporting more and more software engineering activities. Unfortunately, these methods are not normalized, their terminology is not sufficiently unified[8] and not integrated with a software engineering methodology. They are just limited to bio-inspired algorithms and applications of those algorithms[9,10]. Even if these bio-inspired techniques were normalized and integrated in a conventional software engineering methodology, this integration might be inappropriate and incompatible. A bio-inspired methodology, integrating bio-inspired methods might be better and consequently, bio-inspired software engineering might be required.

Despite the huge study which is increasingly carried out in bio-inspired computing field, bio-inspired software engineering was not approached. All these studies might be generally classified into two categories[10]: Algorithms (design, improvements and analysis) and applications of algorithms. These algorithms are inspired from bio-processes which are produced by bio-production engineering. Focusing the study interest only on a product and omitting how this product was produced is a restrictive research method. For this reason, one of the main problems these algorithms have is the lack of a universal platform and of a proper methodology unifying, abstracting and integrating them harmoniously.

Another kind of bio-inspired study deal with software modeling. Mili and El Meslati[8] stated the proliferation of bio-inspired systems and the lack of common agreement on definitions and concepts. They present three different views for biological systems: The phylogenetic (mutation), the ontogenetic (growth) and the epigenetic (learning) without any insinuation to the bio-engineering behind these views. Bakhouya and Gaber[9] and Krupitzer et al.[11] stated the need of adaptive systems to bio-inspired approaches which may be adapted from existing ones. Others studies[12-17] present bio-inspired modeling approaches to specific activities in software engineering (design, implementation and evolution). They do not deal with the whole software engineering life cycle.

It is noteworthy that modern emergent systems like software product lines[18], self adaptive systems[11] and continuous software engineering[3] are mainly based on bio-inspired concepts: Variability, automaticity, dynamicity, autonomy and self control, although they do not implement them with bio-inspired approaches. This suggests that the technology of emergent software systems converges to bio-inspired software engineering.

It is easy to state from the present relevant studies, that bio-inspired technology is currently concerned with particular cases. A generalization of this technology from cases to patterns of theses cases (inspired from bio engineering) will increase the reuse of software artifacts, reduce the cost and increase the quality of software products.

## MATERIALS AND METHODS

Some bio-inspired software engineering basic concepts are abstracted and synthesized from recent and relevant study perspectives. Generally, these concepts are important requirements of emerging systems like software product lines[18], self adaptive systems[11], continuous software engineering[3], etc. They include bio-inspired features and methodology.

**Features:** A bio-inspired software engineering must support the following required features by new trends in software systems:

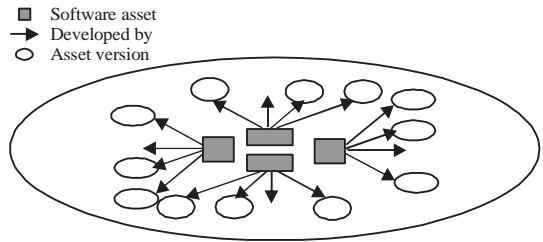

Fig. 1: Software asset variability

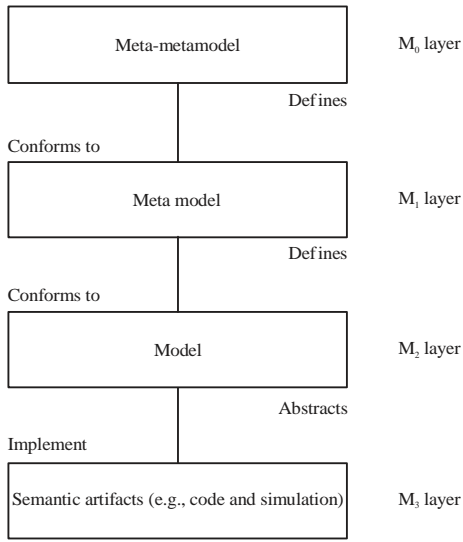

Fig. 2: Meta-modeling patterns hierarchy

- **Variability:** The variability is the stone bed concept of bio-inspired software engineering. It means the availability of any software asset (function, process, methodology, etc.) on multi versions. Figure 1 depicts such variability. Variability introduces relations (implication, exclusion, etc.) between assets versions. A selection mechanism is needed for creating a coherent software product. This variability is well supported in software product lines and in software configuration management systems. While, it is in conventional software engineering limited to some elementary assets (functions and data), it is comprehensive (for all assets) in bio-inspired software engineering (methodologies, processes, functions, etc.)
- **Individuality:** This feature allows software to automate (by self-decisions) adaptation to new environments and requirements, evolution and maintenance, mutation and learning. This automation occurs statically and/or dynamically. Several implementation mechanisms are emerging in self adaptive systems
- **Dynamicity:** This feature allows software, during its run time to take decisions and carry them out. Several implementation mechanisms are emerging in self adaptive systems and software product lines. While, this dynamicity is limited in conventional software engineering, it is comprehensive in bio-inspired software engineering (adaptation, evolution, mutation and learning)
- **Continuity:** This feature allows self propagation of effects of any operation on an asset on all implied ones. This is like pipelines. Some implementations are suggested in continuous software engineering for specific needs
- **Forward:** This feature allows the production methodology to go always forward and never returns back. In conventional software engineering, the control of a methodology allows going back from one activity to another, because time dimension is not captured. Whereas, in bio-inspired software engineering, time dimension is very important and nothing may return back to the past
- **Meta engineering:** In bio-engineering, production systems are based on variability in engineering patterns. Humans and birds etc., are bio-engineering production systems having a common pattern which is instantiated with different versions of some pattern components. This leads to engineering patterns hierarchy, ending (at leaves) by specific engineering. This is a generalization of meta modeling in conventional software engineering. Figure 2 depicts meta-modeling patterns hierarchy

**Methodology:** A bio-inspired software engineering methodology must support at least all the previous software required features. The dominant one is its strong dependency on time dimension, letting it to be only forward. In fact, any bio-engineering product springs in an elementary state, grows up through its temporal trajectory and passes away at the end, never returns back to its past. This is a determinant difference with the conventional software engineering where all its methodologies allow reverse engineering and reengineering of software products. In bio-inspired software engineering, reengineering is part of forward engineering (Fig. 3).

While, in conventional software engineering, a software product evolves on a single dimension, sequencing mutation, growth and learning in bio-inspired software engineering, a software product evolves in a parallel way on three distinct dimensions: Phylogenetic (P: Mutation), ontogenetic (O: Growth) and epigenetic (E: Learning) naturally through the time dimension (Fig. 4).

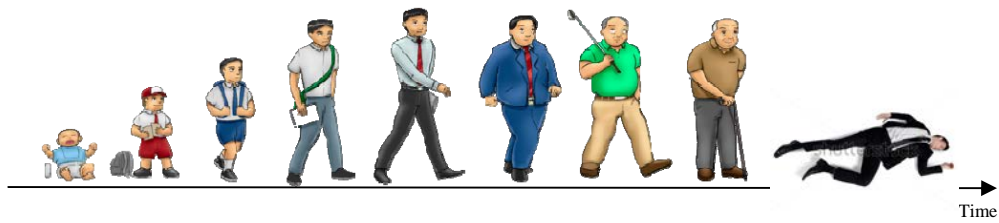

Fig. 3: Temporal trajectory of product in bio-engineering

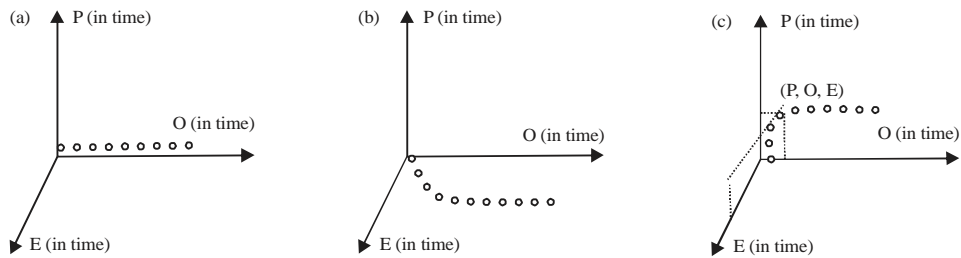

Fig. 4(a-c): (a) Only growing, (b) Growing and learning and (c) Growing, learning and mutation

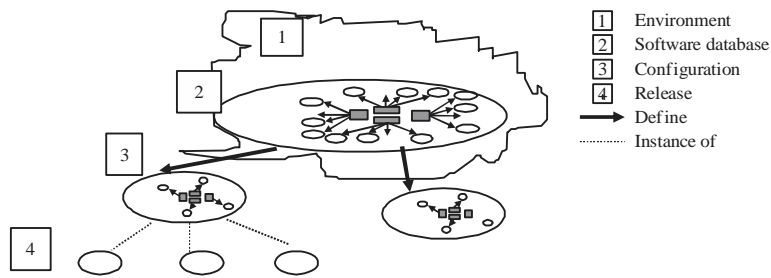

Fig. 5: Software rise

A bio-inspired software engineering might be based on the following four principles.

**Principle 1 (software rise):** Each software system is a release (instance) generated from a configuration defined on a software database (Fig. 5).

This principle suggests the first three steps in bio-inspired software engineering:

- **Software database definition:** This huge database might be specialized in a specific business domain or general according to patterns abstraction hierarchy supported power. It contains all software assets in multiple versions, their relations and their coherence control processes. Thus, the methodology supporting product evolution along with its parallel sub methodologies supporting mutation, growth and learning are also included

- **Configurations definitions:** A configuration is a program selecting versions of assets (one version for any one) composing a software product when executed. These configurations are parts of the software database with their variations, relations and control processes

- **Product instantiating (release):** A product might be instantiated from a configuration. It holds all the assets selected versions by its configuration. The product is then at its first state (age = 0). According to its environment and to its owned growing, learning and mutating methodologies the rising product evolves in the time through its evolution space defined by the three dimensions (P, O and E) tracing its trajectory. Like that at any time, it is possible to answer the following three questions: Where this product is coming from (its past)? What is it actually (its present)? And what is its probable evolution (its probable future)? This provides a rich and precise semantics of software product allowing its automatic understandability and then its self evolution

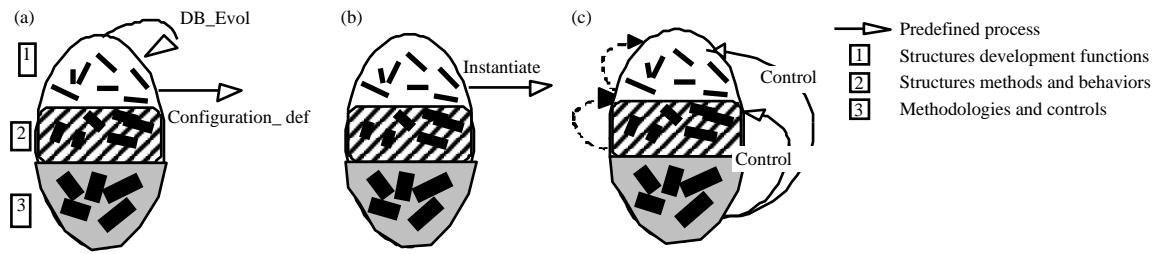

Fig. 6(a-c): Operative nature of (a) Software database, (b) Configuration and (c) Instance

Naturally, each software database modification engenders a mutation, which will only affect new products. An aged product might mutate through its dynamic evolution according to the environment effect.

**Principle 2 (Software components nature):** The nature of software database, configuration and release is completely operative (Fig. 6).

This principle suggests the completely functional nature of any component in the software database, configuration and instance. Structures are built by structures building functions, they operate and behave through operating and behavioural methods and the evolution is supported and controlled by methodologies and control processes.

**Principle 3 (Instance incremental evolution):** The structures, functions and behaviors of an instance are developed incrementally in the time and through the (P, O and E) space.

The encoded evolution methodology along with its sub methodologies (Mutation, growth and learning) ensures that structures, their methods and their behaviors are developed incrementally by time.

**Principle 4 (Control automation):** Software database, configuration definition, product instantiation, growth, evolution and learning are controlled automatically.

The methodology and controls part of the software database ensure the correct evolution of any instance as it is predefined (statically) and according to the environment effect (dynamically). This allows self control of the evolution trajectory.

**RESULTS AND DISCUSSION**

There are no really close studies to compare with this study. A real evaluation necessitates a large adoption of bio-inspired software engineering in the industry, which is far to happen. Then, some aspects were evaluated by specific applications.

**Software database modeling for aspect-oriented systems:** Ghoul[13], presented a bio-inspired aspect-oriented software database modeling approach along with its supporting methodology. This approach has enriched the aspect-oriented paradigm and supporting methodologies with several useful concepts and processes.

**Software database modeling for software product lines:** Younis et al.[17] and Younis and Ghoul[19], presented a software database bio-inspired model based on features diagrams and supporting variability. A methodology supporting this modeling, configuration definition and product instantiation is also provided. The bio-inspired approach has lead to enhancements on software product lines modeling methodology: Features diagrams, configuration definition and product instantiation.

**Software database modeling for object-oriented systems:** Hamouda et al.[14], presented a bio-inspired object-oriented software database modeling approach along with its supporting methodology. This approach has enriched the object-oriented paradigm and supporting methodologies with several useful concepts and processes. Mainly the relation is a implementing inheritance was devalued, whereas the value of the relation composed-by between classes has been increased.

**Self adaptive software methodology:** Naffar and Ghoul[15], present a bio-inspired methodology, included in a bio-inspired software database that supports self adaptive systems growth and mutation. This approach has enriched self adaptive systems engineering with a supporting methodology and several useful concepts and processes. Mainly this study is an example of a definition of growing and mutating methodologies inside a bio-inspired database and their effectiveness.

**Meta engineering:** Al Sultan and Ghoul[16] presented a bio-inspired software meta-modeling approach along with its supporting methodology. The obtained result states clearly the suitability of features diagram formalism instead of UML for this kind of meta-modeling and identifies UML possible enhancement that may generalize it to support software variability meta-modeling.

These applications have not only proved the ease of the feasibility of some important aspects of bio-inspired software engineering (software database, configuration, instance, evolution methodology, mutation methodology, growing methodology, self control and meta engineering etc.) but have also induced valuable enhancements in conventional software engineering.

## CONCLUSION

This study has outlined motivations to bio-inspired software engineering, proposed some fundamental features and methodological principles and ended by an evaluation of such bio-inspired software engineering basics with applications in its software database modeling methodologies, its mutation and growing methodologies and in its meta-engineering. The obtained results have demonstrated the relatively ease of bio-inspired software engineering basics implementation and have a valuable impact on conventional software engineering. However, several important challenges are to be faced.

The software database is too large and complex. Its suitable model will require more efforts and experimentations. These experimentations have proved the non relevancy of object-oriented and aspect-oriented modeling paradigms without important enhancements.

The self, dynamic and continuous (adaptation, mutation and learning) are so far to be understood and mastered in conventional software engineering. Because these features are inherent to nature, their study in the context of bio-inspired software engineering will lead to valuable end relatively simple solutions.

The meta-engineering patterns development by generalization of conventional meta-modeling techniques will lead to improve software quality and reduce considerably its complexity and understandability.

The actual computer architecture might be enhanced in order to efficiently support such powerful concepts, inspired from bio-engineering. Van Newman computer model was designed for computations below this natural level.

The evaluation at an industrial scale should be a dominant challenge, because it will be the decision maker of the acceptance or reject of such engineering.

## ACKNOWLEDGMENT

This study is sponsored from 12/12/2013 to 12/12/2016 by Philadelphia University, through the research project "Bio-inspired systems variability modeling" at the bio-inspired system modeling research laboratory.